\documentclass[
prd,
reprint,
floatfix,
aps,
longbibliography,
superscriptaddress,
nofootinbib
]{revtex4-2}

\usepackage{amsmath,graphicx,amssymb,xcolor,booktabs,braket,multirow,cancel,enumerate,enumitem,bbold}
\usepackage{mathrsfs}
\usepackage[colorlinks=true, allcolors=blue]{hyperref}
\usepackage[normalem]{ulem}
\usepackage{soul}
\usepackage{slashed}

\setstcolor{red}

\begin{document}

\title{
    A quality-coupling relation in chiral \texorpdfstring{$U(1)_{B-L}$}{U1BL} axion model
    }

\author{Yu-Cheng Qiu}
\email{ethan.qiu@cityu.edu.hk}
\affiliation{Department of Physics, City University of Hong Kong, Kowloon, Hong Kong SAR, China}

\date{\today}

\begin{abstract}
The quality of the QCD axion can be protected by a gauge symmetry.
This work considers a gauged chiral $U(1)_{B-L}$.
The vacuum expectation values of two complex scalars break $U(1)_{B-L}$. One linear combination of their two phase degrees of freedom is eaten by the $U(1)_{B-L}$ gauge boson, while the other becomes the QCD axion.
Since the PQ symmetry is accidental, there is no ambiguity in the PQ charge assignment.
High axion quality implies a large anomalous coupling between the axion and the $U(1)_{B-L}$ gauge boson (or gauged Majoron).
A `quality floor', namely a minimal coupling determined by the axion quality, arises for the coupling between the axion and SM fermions,
extending the parameter space of QCD axions toward improved testability.
\end{abstract}

\maketitle

\section{Introduction}

The strong CP problem remains one of the central naturalness puzzles of the Standard Model.
Quantum chromodynamics (QCD) contains a CP-violating parameter $\bar \theta$, which is constrained by measurements of the neutron electric dipole moment to satisfy $\bar \theta < 10^{-10}$~\cite{Smith:1957ht,Abel:2020pzs}.
There is no known anthropic reason for this parameter to be so small.
Therefore, a theoretical explanation is needed.

The QCD axion provides a viable solution by introducing a global anomalous Peccei-Quinn (PQ) symmetry~\cite{Peccei:1977hh,Peccei:1977ur}, whose low-energy Nambu-Goldstone boson is the QCD axion~\cite{Weinberg:1977ma,Wilczek:1977pj,Kim:1979if,Shifman:1979if,Dine:1981rt,Zhitnitsky:1980tq}.
Through the anomaly, the axion couples to the gluon field and absorbs the QCD CP-violating vacuum parameter $\bar \theta$ via a shift, effectively promoting it to a dynamical field $\bar \theta(x)$.
Instantons generate a potential for $\bar \theta(x)$ that dynamically drives $\langle \bar \theta(x) \rangle \to 0$, resolving the strong CP problem.
However, quantum gravity can explicitly break the global PQ symmetry~\cite{Banks:2010zn}, contributing an additional term to the potential of $\bar \theta(x)$ and shifting its vacuum value.
This spoils the solution and reintroduces the strong CP problem.
This is the axion quality problem.

Several approaches have been proposed to suppress PQ-breaking operators and resolve the quality problem. One is to make the PQ symmetry accidental~\cite{Georgi:1981pu,Redi:2016esr,DiLuzio:2017pfr,Duerr:2017amf,Fukuda:2017ylt,Bonnefoy:2018ibr,Lillard:2018fdt,Lee:2018yak,Fukuda:2018oco,Ibe:2018hir,Gavela:2018paw,Ardu:2020qmo,Choi:2020vgb,Darme:2021cxx,Contino:2021ayn,Choi:2022fha,Qiu:2023los,Sheng:2025sou}.
Another approach introduces additional mechanisms, including extra-dimensional setups, to modify the renormalization group (RG) running of the breaking operator and suppress it at low energies~\cite{Izawa:2002qk,Choi:2003wr,Izawa:2004bi,Flacke:2006ad,Cox:2019rro,Bonnefoy:2020llz,Nakai:2021nyf,Nakagawa:2023shi,Nakagawa:2024kcb,Craig:2024dnl,Reece:2025thc,Choi:2025lkg,Choi:2026kxu,Csaki:2026qjl}.
A sufficiently heavy axion can also be automatically free from the quality problem~\cite{Tye:1981zy,Hook:2019qoh,Liu:2021wap,Girmohanta:2024nyf,Murayama:2026ioh}.
Another possibility is that there is no quality problem at all, since little is known about quantum gravity~\cite{Dvali:2022fdv,Burgess:2023ifd,Catinari:2024zon}.
The existence of these competing possibilities motivates a deeper investigation of the problem and of how the different scenarios might be tested.
This work focuses on the first approach.

In the Wilsonian effective field theory (EFT) approach,
one should write down all possible operators, including nonrenormalizable operators, that are consistent with the relevant symmetry requirements.~\footnote{A systematic analysis of the quality problem in EFT is provided in Ref.~\cite{Cheek:2026dvu}.}
If the PQ symmetry is accidental, for example because it is embedded in a gauge symmetry that cannot be broken by gravity, then PQ-breaking operators are suppressed by higher powers and the axion quality is protected.
For example, Ref.~\cite{Qiu:2023los} proposes a chiral $U(1)$ gauge theory that protects the PQ symmetry.
An additional gauged $U(1)$ is also well motivated in cosmological contexts, where it can give rise to rich phenomena, such as complex topological defects~\cite{Binetruy:1998mn,Hiramatsu:2019tua,Niu:2023khv,Eto:2024hwn,Hua:2026mgn}.
Interestingly, another well-studied beyond SM scenario is a gauged $U(1)_{B-L}$, where $B$($L$) refers to baryon (lepton) number.
To gauge $U(1)_{B-L}$, one must introduce three right-handed neutrinos to cancel the gravitational anomaly. This connects to the construction to neutrino masses~\cite{Mohapatra:1979ia,Mohapatra:1980qe}, making leptogenesis~\cite{Fukugita:1986hr,Buchmuller:2005eh} possible.
It has recently been realized that the $U(1)_{B-L}$ gauge boson can also be an ultralight dark-matter candidate (f\'eeton)~\cite{Lin:2022xbu,Sheng:2023iup}.

This work proposes that an extended chiral $U(1)_{B-L}$ gauge theory can be used to protect the QCD axion quality under the QWY framework~\cite{Qiu:2023los}.~\footnote{QWY refers to the initials of the authors' surnames.}
The quality of the axion depends on the number of additional heavy quarks and on the anomaly-free charge assignments.
High axion quality usually requires a large number of additional fermions, which decouple from the SM in the low-energy limit.
However, because they carry anomalous PQ charges, anomaly matching implies phenomenological consequences in the IR.
This work examines the resulting modifications to the effective QCD-axion Lagrangian.
Specifically, a large number of additional fermions carrying $U(1)_{B-L}$ charges gives rise to a large anomalous coupling between the axion and the $U(1)_{B-L}$ gauge boson, which sets a lower bound on the coupling between the axion and SM fermions.
The axion quality may therefore be testable if the axion is discovered.

\section{The chiral \texorpdfstring{$U(1)_{B-L}$}{U1BL} model}
\label{sec:chiral-model}

Here I normalize the $B-L$ charge in units of $1/3$, which means that the Standard Model quarks carry $U(1)_{B-L}$ charge $1$ and leptons carry $-3$.
The gauge anomaly cancels within the SM if three right-handed neutrinos are added.
To realize a high-quality QCD axion, this framework introduces two complex scalars $\phi_{1,2}$ that carry $U(1)_{B-L}$ charge $q_{1,2}$,
and $N=k+l$ pairs of chiral quarks that are also charged under $U(1)_{B-L}$,
of which $k$ ($l$) pairs couple to $\phi_1$ ($\phi_2$) through Yukawa interactions.
One of the $\phi_i$ can give mass to the right-handed neutrinos.
Here I keep the discussion general.
The essence of this theory is
\begin{equation}
-\mathcal L \supset \sum_{j=1}^k y_{\chi,j} \phi_1 \bar \chi_j \chi_j + \sum_{j=1}^l y_{\psi,j} \phi_2 \bar \psi_j \psi_j + V(\phi_1, \phi_2)\;,
\end{equation}
where $\chi,\psi \in (\mathbf {3}, 1, 0)$ and $\bar \chi, \bar \psi \in (\mathbf 3^*,1,0)$ under the SM $SU(3)_{\rm c} \times SU(2)_{\rm L} \times U(1)_{\rm Y}$ gauge group.

To avoid cross-Yukawa terms between the $k$ and $l$ pairs,
I choose the $U(1)_{B-L}$ charges of the additional quark pairs such that
\begin{align}
	C_{\bar\chi,j} + C_{\chi,j} & = - q_1 \;, & C_{\bar \psi,j} + C_{\psi,j} & = -q_2\;, \nonumber\\
	C_{\bar \chi, i} + C_{\psi, j} & \neq \pm q_{1,2}\;, & C_{\bar \psi, i} + C_{\chi, j} &\neq \pm q_{1,2}\;.
	\label{eq:charge_assignment}
\end{align}
Cross terms can be allowed for particular charge assignments.
However, the calculation would then be case dependent and cumbersome.
The no-cross-coupling conditions are imposed for simplicity.
Cancellation of the gauge and gravitational anomalies requires that
\begin{subequations}
\begin{align}
	\sum_{j=1}^k \left( C_{\bar \chi,j} + C_{\chi,j} \right) + \sum_{j=1}^l \left( C_{\bar \psi,j} + C_{\psi,j} \right)       & = 0 \;, \label{eq:AC_1} \\
	\sum_{j=1}^k \left( C_{\bar\chi,j}^3 + C_{\chi,j}^3 \right) + \sum_{j=1}^l \left( C_{\bar\psi,j}^3 + C_{\psi,j}^3 \right) & = 0 \;.\label{eq:AC_2}
\end{align}
\end{subequations}
The first condition~\eqref{eq:AC_1}, together with the charge assignments, gives the relation~\cite{Qiu:2023los},
\begin{equation}
	-\frac{q_1}{q_2}=\frac{l}{k} = \frac{m}{n}\;, \quad N_{\rm DW} = \frac{l}{m} = \frac{k}{n}\;.
	\label{eq:QWY_condition}
\end{equation}
where the integers $m$ and $n$ are coprime.
$N_{\rm DW}$ is the domain wall number.~\footnote{The derivation of $N_{\rm DW}$ is shown in the appendix of Ref.~\cite{Qiu:2023los}.}
This relation connects the axion quality to the number of additional quarks introduced.
The second condition~\eqref{eq:AC_2} is used to find the solutions numerically.~\footnote{Explicit solutions can be found using the methods developed in Ref.~\cite{Costa:2019zzy}.}
The two complex scalars give rise to two global $U(1)$s. One linear combination is gauged as $U(1)_{B-L}$, while the orthogonal combination becomes the accidental PQ symmetry $U(1)_a$.
Suppose that $V(\phi_1, \phi_2)$ gives the complex scalars vacuum expectation values (VEVs) $\langle \phi_i \rangle = v_i/\sqrt{2}$, spontaneously breaking $U(1)_{B-L}$ (together with $U(1)_a$).
Focusing on their phase directions, $\phi_i \to v_i e^{i \theta_i}/\sqrt{2}$,
the corresponding mixing is
\begin{align}
	\begin{pmatrix}
		v_1 \theta_1 \\
		v_2 \theta_2
	\end{pmatrix}
	=
	\begin{pmatrix}
		\cos \beta & - \sin \beta \\
		\sin \beta & \cos \beta
	\end{pmatrix}
	\begin{pmatrix}
		F_a \theta_a \\
		v_b \theta_b
	\end{pmatrix} \;,
	\label{eq:mixing}
\end{align}
where $\tan \beta = m v_1/n v_2$.
Then, the kinetic terms $\left| D_\mu \phi_1 \right|^2 + \left| D_\mu \phi_2 \right|^2$ become
\begin{equation}
	 \frac{1}{2}[\partial_\mu (F_a \theta_a)]^2 + \frac{1}{2} m_B^2 \left[B_\mu -\frac{1}{m_B} \partial_\mu (v_b \theta_b) \right]^2\;,
\end{equation}
where $m_B = g\sqrt{q_1^2 v_1^2 + q_2^2 v_2^2}$.
Here $\theta_b$ becomes the longitudinal mode of $B_\mu$ and $\theta_a$ is the axion.
The field range and decay constant $F_a$ of the axion are
\begin{align}
	a & \equiv F_a \theta_a \in \left( 0, \frac{2\pi v_1v_2}{\sqrt{m^2 v_1^2 + n^2 v_2^2}} \right] \;, \\
	F_a & = \frac{v_1v_2}{\sqrt{m^2 v_1^2 + n^2 v_2^2}}\;. \nonumber
\end{align}
The shift of $\theta_a$ is anomalous.
At low energies, anomaly matching requires the term
\begin{equation}
\left( \bar \theta + N_{\rm DW} \frac{a}{F_a} \right) \frac{g_s^2}{32\pi^2} G_{\mu\nu}^A \tilde G^{A\mu\nu}\;,
\end{equation}
where $G_{\mu\nu}^A$ is the gluon field strength,
and $\tilde G^{A\mu\nu} = \epsilon^{\mu\nu\sigma \rho} G_{\sigma \rho}^A/2$.
This resolves the strong CP problem once the axion is stabilized at $\langle a \rangle \to \bar \theta F_a /N_{\rm DW}$.
The quality of this axion is determined by operators that explicitly breaks the shift symmetry, which in this theory is
\begin{equation}
	\mathcal O_a = \frac{1}{n! m!} \frac{\phi_1^n \phi_2^m}{M_{\rm Pl}^{N/N_{\rm DW} - 4}} + \cdots \;.
\end{equation}
The higher-order terms are neglected from now on.
Note that $\mathcal O_a$ is invariant under the $U(1)_{B-L}$ transformation thus allowed in the EFT.
However, the shift symmetry is broken by $\mathcal O_a$, which introduces an additional contribution to the axion potential,
\begin{align}
	c_a\mathcal O_a \to \delta V_a & = \frac{1}{n! m!} \left( \frac{m^2+n^2}{2} \right)^{N/2N_{\rm DW}}  \label{eq:quality} \\
	& \qquad \times \left(\frac{F_a}{M_{\rm Pl}} \right)^{N/N_{\rm DW}} M_{\rm Pl}^4 \cos\left(\frac{a}{F_a} + \delta \right) \;, \nonumber
\end{align}
where $\delta$ is an arbitrary phase arising from the Wilson coefficient $c_a$, and we assume $v_1 = v_2$ for simplicity.
For large $m$ and $n$ (thus large $N/N_{\rm DW}$), this operator is sufficiently suppressed.

In addition to its anomalous coupling to gluons, the axion also couples to the $U(1)_{B-L}$ gauge field through
\begin{equation}
\mathcal  C_{B-L} \times \frac{g_{B-L}^2}{32\pi^2} \frac{a}{F_a} B_{\mu\nu} \tilde B^{\mu\nu}\;,
\end{equation}
where $g_{B-L}$ is the $U(1)_{B-L}$ gauge coupling and the anomaly coefficient is
\begin{align}
\mathcal C_{B-L} & = N_{\rm c}\frac{n v_2^2 c_1 + m v_1^2 c_2}{m^2 v_1^2 + n^2 v_2^2} \;, \label{eq:C_BL}\\
c_1 & = \frac{1}{3} \sum_{j=1}^k \left( C_{\chi,j}^2 + C_{\bar \chi,j}^2 - C_{\chi,j} C_{\bar\chi,j} \right) \;, \nonumber\\
c_2 & = \frac{1}{3} \sum_{j=1}^l \left( C_{\psi,j}^2 + C_{\bar \psi,j}^2 - C_{\psi,j} C_{\bar\psi,j} \right)\;. \nonumber
\end{align}
A detailed calculation is provided in the Appendix.
Here the PQ symmetry is embedded in $U(1)_{B-L}$, which means that there is no ambiguity in the PQ charge assignment.
Therefore, this anomalous coupling is entirely fixed by the $U(1)_{B-L}$ charge assignment (UV data).

\section{Anomalous coupling}

For a high-quality QCD axion, sufficiently large values of $k$ and $l$ are required according to Eq.~\eqref{eq:quality}.
This implies a large anomalous coupling $\mathcal C_{B-L}$.
In a simplified case, I demonstrate the direct connection between the axion quality and $\mathcal C_{B-L}$ by deriving a lower bound.

\begin{figure}
\centering
\includegraphics[width=8cm]{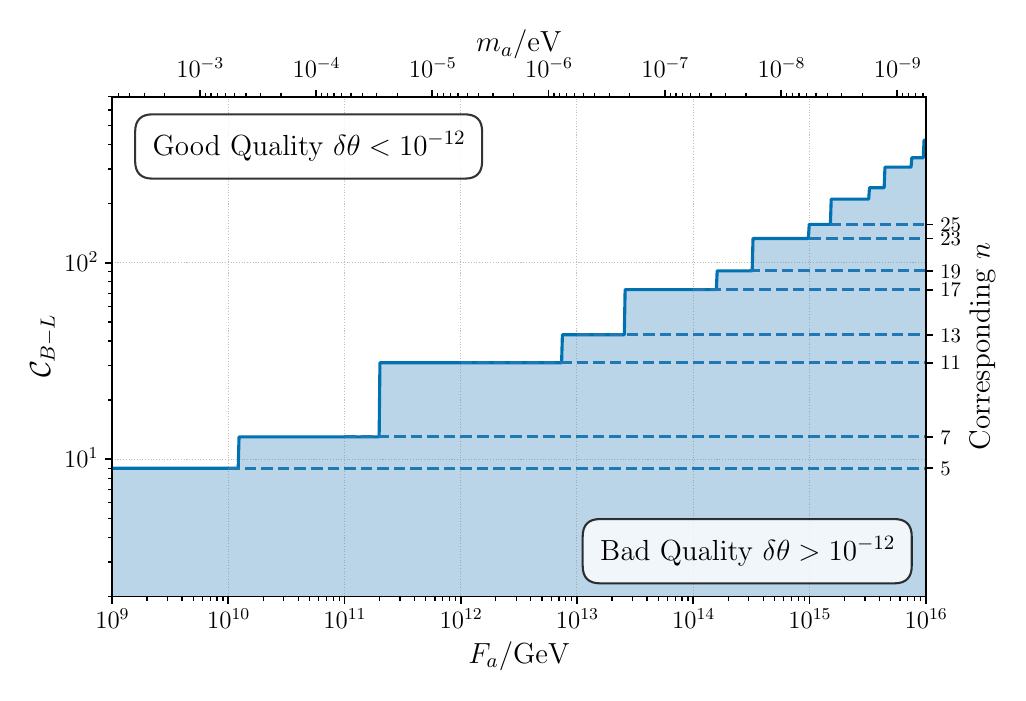}
\caption{Relation between the decay constant $F_a$ and anomalous coupling $\mathcal{C}_{B-L}$.
Here $m=6$ is fixed. The available choices of $n$ are $\{1,5,7,11,13,17,19,\cdots\}$.
The values of $n$ shown on the right correspond to $(\mathcal{C}_{B-L})_{\rm min}$ and are not in one-to-one correspondence with the actual values of $\mathcal{C}_{B-L}$.
The actual anomalous coupling is determined by the explicit charge assignments of all heavy quarks, namely the complete UV data.}
\label{fig:C_Fa}
\end{figure}

Here consider two scalars with the same VEV, $v_1 = v_2$~\footnote{The cosmological scenario with $v_1 \neq v_2$ is discussed in Ref.~\cite{Zhou:2026amj}.}, and assign
\begin{equation}
q_1 = m = l \;,\quad - q_2 = n = k\;.
\label{eq:simple}
\end{equation}
So, $N_{\rm DW} = 1$.
Take $N_c = 3$ for QCD, and parametrize
\begin{align}
C_{\chi,j} & = -\frac{q_1+ a_j}{2}\;, & C_{\bar \chi,j} & = -\frac{q_1 - a_j}{2}\;, \nonumber \\
C_{\psi,j} & = -\frac{q_2 + b_j}{2}\;, & C_{\bar \psi,j} & = -\frac{q_2 -b_j}{2}\;,
\end{align}
where $a_j$ and $b_j$ are integers satisfying
\begin{align}
    a_j & \sim m \mod 2 \;, \nonumber\\
    b_j & \sim n \mod 2 \;. \label{eq:a_jb_j}
\end{align}
The cubic anomaly-cancellation condition~\eqref{eq:AC_2} becomes
\begin{equation}
nm\left( m^2 - n^2 \right) + 3 m A - 3 n B = 0\;,
\label{eq:cubic}
\end{equation}
where $A= a_1^2 + a_2^2 + \cdots + a_n^2$ and $B= b_1^2 + b_2^2 + \cdots + b_m^2$.
The anomaly coefficient~\eqref{eq:C_BL} is then
\begin{equation}
\mathcal C_{B-L} = \frac{1}{4}\left(m^2 + \frac{3A}{n} \right) = \frac{1}{4} \left(n^2 + \frac{3B}{m}\right)\;.
\end{equation}
By definition, $A,B\geq 0$, so $\mathcal C_{B-L}$ must be nonzero.
Since $m$ and $n$ are coprime, they cannot both be even.
Three cases then follow.
\begin{enumerate}[leftmargin=12pt]
\item Assume that $n$ is even and $m$ is odd. According to Eq.~\eqref{eq:a_jb_j}, one has $(a_j^2)_{\rm min} = 1$ and $(b_j^2)_{\rm min} = 0$,
which leads to $B\geq 0$ and $A\geq n$. Meanwhile, the cubic equation~\eqref{eq:cubic} indicates that $3nB = nm(m^2 -n^2) + 3mA$.
Therefore, $B\geq0$ implies $A \geq n(n^2 - m^2)/3$.
Thus, the allowed range of $A$ is $A\geq \max(n, n(n^2- m^2)/3)$.
This gives
\begin{equation}
\mathcal C_{B-L} \geq \max\left(\frac{n^2}{4},\, \frac{m^2 + 3}{4} \right)\;.
\label{eq:C_min}
\end{equation}
Using the corresponding range for $B$ yields the same result.

\item If $m$ is even and $n$ is odd, then $A\geq 0$ and $B\geq m$.
An analogous analysis yields the result obtained by interchanging $n\leftrightarrow m$ in Eq.~\eqref{eq:C_min}.

\item If $m$ and $n$ are both odd, then $A\geq n$ and $B\geq m$.
Using the cubic equation~\eqref{eq:cubic}, one obtains
\begin{equation}
\mathcal C_{B-L} \geq \max\left(\frac{m^2 + 3}{4},\, \frac{n^2 + 3}{4} \right)\;.
\end{equation}
\end{enumerate}
The analysis above does not fully account for the integer nature of the solutions $a_j$ and $b_j$.
The actual minimum can therefore be larger than the bounds found here.
Assuming that $\phi_1$ is the Majoron that gives mass to the right-handed neutrinos (with charge $-3$ under the $1/3$ normalization), one can fix $m = 6$.
Then the choice of $n$ and $F_a$ determines the quality and $\mathcal C_{B-L}$.
The shift $\delta \bar \theta$ is induced by $\delta V_a$ in Eq.~\eqref{eq:quality} and must satisfy $\delta \bar \theta < 10^{-10}$.
Explicitly, the quality requirement is
\begin{align}
& \frac{1}{n! m!} \left( \frac{m^2+n^2}{2} \right)^{(m+n)/2}  \\
&\qquad \times \left(\frac{F_a}{M_{\rm Pl}} \right)^{m+n} \frac{M_{\rm Pl}^4}{m_\pi^2 f_\pi^2} \frac{(1+z)^2}{z} \leq 10^{-10}\;, \nonumber
\end{align}
where $m_\pi =135$~MeV, $f_\pi = 92$~MeV, and $z = m_u/m_d = 0.48$~\cite{GrillidiCortona:2015jxo}.
The relation between $\mathcal C_{B-L}$ and $F_a$ is shown in Fig.~\ref{fig:C_Fa}.
For a fixed $F_a$ and chosen $m$, better axion quality requires a larger value of $n$, which sets a higher $(\mathcal C_{B-L})_{\rm min}$.
Thus,
small $\mathcal C_{B-L}$ is excluded by the quality requirement $\bar \theta <10^{-10}$.

If one measures axion quality by the smallness of $\delta \theta$, then large $\mathcal C_{B-L}$ can be regarded as the `price' that must be paid.
For example, at $F_a = 10^{12}$~GeV, achieving $\delta \theta < 10^{-10}$ requires $n\geq 11$ (assuming $m=6$).
One can assign $(a_1,a_2,\cdots, a_{11}) = (0,0,0,0,0,0,2,2,2,-12,14)$ and $(b_1,b_2,\cdots,b_{6}) = (-1,-1,1,1,3,3)$,
which satisfies the cubic equation~\eqref{eq:cubic}, cancelling the anomaly.
This gives $\mathcal C_{B-L} = 33$.
Imposing the no-cross Yukawa condition~\eqref{eq:charge_assignment} would not alter this theoretical lower bound, which is only necessary when looking for explicit charge assignments.

\section{Axion-fermion coupling}

Even if the axion has no direct coupling to SM fermions, denoted as $f$, at the
PQ scale, an effective axion-fermion interaction is
radiatively generated by the anomalous $\mathcal C_{B-L}$ coupling.
Consider
\begin{align}
\mathcal L & \supset {}
N_{\rm DW}\frac{g_s^2}{32\pi^2}
\frac{a}{F_a} G_{\mu\nu}^A \tilde G^{A\mu\nu}
+
\mathcal C_{B-L}\frac{g_{B-L}^2}{32\pi^2}
\frac{a}{F_a}B_{\mu\nu} \tilde B^{\mu\nu}
\nonumber\\
&\quad + g_{B-L} B_\mu \sum_f q_f \bar{f} \gamma^\mu f
+\frac{1}{2}m_B^2B_\mu B^\mu .
\end{align}
Here the charge is $q_f = 1$ ($-3$) for SM quarks (leptons).

The effective pseudoscalar interaction
is parametrized as
\begin{equation}
\mathcal L_{\rm eff}
\supset
i\sum_f g_{aff} a\bar f\gamma^5f \;,
\end{equation}
where $g_{aff}$ is dimensionless.
At leading order, the contribution induced by the
massive $B_\mu$ gauge boson is
\begin{equation}
g_{aff}^{(B-L)}
\simeq
\frac{3\alpha_{B-L}^2}{4\pi^2}
\frac{q_f^2 m_f }{F_a}
\mathcal C_{B-L}
\log\frac{v}{m_B}\;,
\end{equation}
where we have assumed that $v_1 = v_2 > m_B > m_f$.
Here $v_i = \sqrt{m^2 + n^2 }F_a$ labels the scale of $U(1)_{B-L}$ breaking.

For a charged lepton $\ell$ below the QCD scale, there
is an additional model-independent electromagnetic contribution~\cite{Bardeen:1977bd,Srednicki:1985xd,Bauer:2020jbp},
\begin{equation}
g_{a\ell\ell}^{(\gamma,\mathrm{QCD})}
\simeq
-\frac{3\alpha_{\rm em}^2}{4\pi^2}
\frac{Q_\ell^2 m_\ell}{F_a}
C_{a\gamma}^{\rm QCD}
\log\frac{\Lambda_{\rm QCD}}{m_\ell}\;,
\end{equation}
where $C_{a\gamma}^{\rm QCD}=1.92(4)$~\cite{GrillidiCortona:2015jxo}.
Thus, taking the electron as an example, one has $g_{aee} \simeq g_{aee}^{(B-L)} + g_{aee}^{(\gamma, {\rm QCD})}$ in low energy, up to finite threshold corrections and convention-dependent
overall signs.
There may exist some cancellation in $|g_{a\ell\ell}|$ between different components.

\begin{figure}
\centering
\includegraphics[width=8cm]{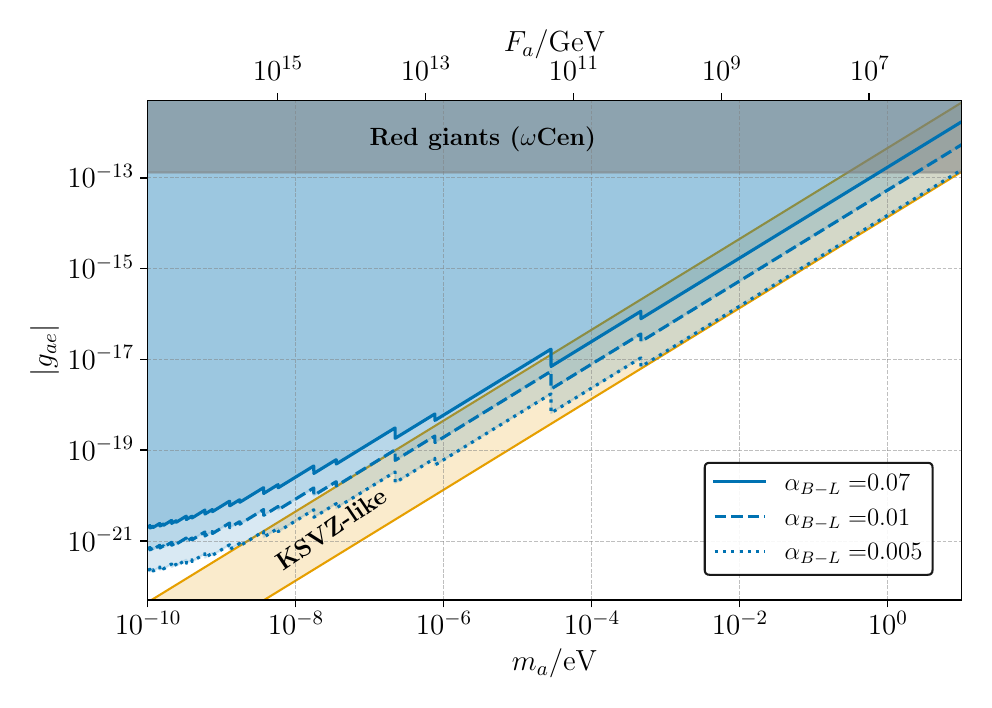}
\caption{Dimensionless axion-electron coupling $|g_{aee}|$ as a function of $m_a$ (or $F_a$). Here $(\mathcal C_{B-L})_{\rm min}$ is calculated under the simplifying assumption in Eq.~\eqref{eq:simple}, with $m=6$. The blue-shaded region denotes the couplings allowed by the quality requirement, and its boundaries correspond to different choices of $\alpha_{B-L}$. The orange band shows the coupling range of the original KSVZ-like axion.
The gray-shaded region is excluded by red giants~\cite{Capozzi:2020cbu}.}
\label{fig:g_aff}
\end{figure}

For a given set of parameters ($m$, $n$, $m_B$, $m_f$, $F_a$, $q_f$), the anomalous coupling $\mathcal C_{B-L}$ is still undetermined unless explicit charge assignments (full UV data) for heavy chiral quarks are given.
However, there is a lower bound, as an IR consequence, on this induced coupling.
It is determined by $(\mathcal C_{B-L})_{\rm min}$ through the quality requirement.
For fixed $F_a$ and $m$, axion quality imposes a lower bound on $n$.
For each specific value of $n$, there is also a lower bound on $\mathcal C_{B-L}$.
Thus, for a fixed $\{F_a,m\}$ and given $\alpha_{B-L}$, the induced coupling $g_{a\ell\ell}^{(B-L)}$ is bounded from below.
As shown in Fig.~\ref{fig:g_aff},
there is a minimum dimensionless coupling $|g_{aee}|$ for a given QCD-axion quality.
The coupling $g_{aee}^{(B-L)}$ can exceed $g_{aee}^{\rm QCD}$ for some choices of $\alpha_{B-L}$ and $m_a$, defining a `quality floor'.
This indicates a broad, previously unexplored parameter space for invisible QCD axions, rather than only a narrow band in KSVZ-like.

\section{Summary and discussion}
\label{sec:discussion}

In this work, I discuss the scenario of using an extended chiral gauge $U(1)_{B-L}$ to protect the quality of the QCD axion under the QWY framework.
High axion quality implies a large anomalous coupling between the axion and the $U(1)_{B-L}$ gauge boson,
which in turn generates an induced axion-fermion coupling.
This extends the parameter space for
invisible QCD axion couplings to SM fermions, with the lower bound directly linked to the axion quality.
The axion quality can therefore be testable,
and the UV data, specifically the heavy chiral quark charge assignments, can have IR consequences in the form of a quality floor.

In another extreme scenario, if $\alpha_{B-L}$ is sufficiently small to make $B_\mu$ stable, the gauge boson can serve as dark matter, known as the f\'eeton~\cite{Lin:2022xbu}.
In that case, the quality floor (minimal coupling) is set by the QCD-induced coupling, $|g_{a\ell\ell}^{(B-L)}|_{\rm min} \ll |g_{a\ell\ell}^{\gamma, {\rm QCD}}|$.
Nevertheless, one could still have $|g_{a\ell\ell}| \gg |g_{a\ell\ell}^{\gamma, {\rm QCD}}|$ for suitable UV data.
The dark portal $aF\tilde B$ is present when kinetic mixing~\cite{Holdom:1985ag} between $U(1)_{B-L}$ and $U(1)_{\rm EM}$ is included.
A detailed study is left for future work.

Since the axion decay constant is determined by the VEVs of the scalars that break $U(1)_{B-L}$,
the axion can couple to neutrinos.
The induced neutral-pion decay $\pi^0 \to \nu \nu$ is strongly suppressed by axion-pion mixing and by the neutrino mass $\Gamma \propto f_\pi^2 m_\nu^2 m_\pi/f_a^4$.

\section*{Acknowledgments}

The author thanks Jie Sheng for valuable comment.
The work is supported by GRF Grants No.~11302824 and No.~11310925, and CityUHK Grants No.~9610645 and No.~7020130.

\appendix

\section{The anomalous coupling}

We begin with a simpler case.
The propagators for a massive Weyl-fermion pair $M \bar\chi \chi + {\rm h.c.}$ are~\cite{Dreiner:2008tw},
\begin{align}
	\langle \chi_\alpha \chi_{\dot\beta}^\dagger \rangle   = \langle \bar\chi_\alpha \bar\chi_{\dot\beta}^\dagger \rangle  & = \frac{ip_\mu \sigma^\mu_{\alpha \dot \beta}}{p^2 - M^2 + i \epsilon} \;,\\
	\langle \chi_\alpha \bar \chi_{\beta} \rangle  & = \frac{i M \epsilon_{\alpha \beta}}{p^2 - M^2 + i \epsilon}\;, \\
	\langle \chi_{\dot\alpha}^\dagger \bar \chi_{\dot\beta}^\dagger \rangle & = \frac{i M \epsilon_{\dot \alpha \dot \beta}}{p^2 - M^2 + i \epsilon}\;,
\end{align}
where $M$ is real for simplicity.
$\chi$ and $\bar\chi$ denote two independent left-handed Weyl fields; the bar on $\bar\chi$ is a field label, not Hermitian conjugation.
Suppose that this fermion pair couples to a Goldstone boson through a mass-flipping term
\begin{equation}
	iM \theta \chi \bar\chi  + {\rm h.c.}\;,
\end{equation}
and that the two fields carry charges $C_{\bar\chi}$ and $C_\chi$, respectively, under the gauge symmetry $U(1)_g$, which is spontaneously broken by $M$.

Consider the process $\theta(q) \to B_\mu(k_1) + B_\nu(k_2)$.
Since $\theta$ flips the chirality, another mass insertion is needed to complete the fermion loop.
There are two possible locations for this insertion: (1) between the two gauge bosons and (2) between one gauge boson and $\theta$.
The amplitude is
\begin{equation}
	\mathcal M(q) =  \varepsilon_\mu(k_1) \varepsilon_\nu(k_2) \left( \mathcal T^{\mu \nu}_{12} + \mathcal T^{\nu\mu}_{21} \right) \;,
\end{equation}
where $\mathcal T^{\mu\nu}_{21} = \mathcal T^{\mu\nu}_{12}(k_1 \leftrightarrow k_2)$, and
\begin{align}
	\mathcal T^{\mu\nu}_{12} & = (i M)^2 (ig)^2 \int \frac{d^4 \ell}{(2\pi)^4} \\
	&\times \frac{N_{12}^{\mu\nu}}{(\ell^2 - M^2) [(\ell + k_1)^2 -M^2][(\ell- k_2)^2 - M^2]} \nonumber
\end{align}
We choose $\ell$ to denote the momentum flowing between the two gauge-boson vertices.
The numerator is obtained by summing the three possible mass-insertion contributions.
These contributions are proportional to $C_\chi^2$, $C_{\bar\chi}^2$, and $C_{\chi}C_{\bar\chi}$.
\begin{align}
N_{12}^{\mu\nu} & = C_{\bar\chi}^2 \,{\rm tr} \left[\sigma^\mu \bar \sigma^\alpha \sigma^\nu \bar \sigma^\beta \right] \ell_\alpha (\ell - k_2)_\beta \\
& \qquad + C_{\chi}^2 \,{\rm tr}\left[\sigma^\alpha \bar \sigma^\mu \sigma^\beta \bar \sigma^\nu \right] (\ell + k_1)_\alpha \ell_\beta \nonumber \\
& \qquad - C_{\bar \chi} C_\chi \, {\rm tr} \left[ \sigma^\alpha \bar \sigma^\mu \sigma^\nu \bar \sigma^\beta \right] (\ell + k_1)_\alpha (\ell - k_2)_\beta \;, \nonumber
\end{align}
where the minus sign in the $C_\chi C_{\bar\chi}$ term arises from the different spinor structures of the gauge couplings, $ i \bar \sigma^\alpha \leftrightarrow - i \sigma^\alpha$.
The spinor identities needed here are
\begin{align}
	{\rm tr}\left[\sigma^\mu \bar \sigma^\alpha \sigma^\nu \bar \sigma^\beta\right]    & = 2i \epsilon^{\mu \alpha \nu \beta} + (\text{even})\;,  \\
	{\rm tr}\left[\bar \sigma^\mu \sigma^\alpha \bar \sigma^\nu \sigma^\beta\right] & = - 2i \epsilon^{\mu \alpha \nu \beta} + (\text{even})\;,
\end{align}
where the parity-even parts are not relevant here.
Thus, the relevant part of the numerator becomes
\begin{align}
N_{12}^{\mu\nu} & = (2i)\epsilon^{\mu\nu \alpha \beta} \bigg[ \ell_\alpha\left(C_\chi^2 k_1 + C_{\bar \chi}^2 k_2 \right)_\beta \\
&\qquad  -  C_\chi C_{\bar \chi} \left( -\ell_\alpha k_{2\beta} + k_{1\alpha} \ell_\beta - k_{1\alpha} k_{2\beta} \right) \bigg]\;. \nonumber
\end{align}
Symmetric terms, such as $\epsilon^{\mu\nu\alpha \beta} \ell_\alpha \ell_\beta = 0$, vanish.
Wick rotation gives
\begin{equation}
\mathcal T^{\mu\nu}_{12} = - 2 g^2 M^2 \epsilon^{\mu\nu\alpha\beta} \int \frac{d^4 \ell_{\rm E}}{(2\pi)^4} \frac{N_{{\rm E},\alpha\beta}}{D_0 D_1 D_2}  + \cdots\;,
\end{equation}
where $D_0  = \ell_{\rm E}^2 + M^2$, $D_1  = (\ell_{\rm E} + p_1)^2 + M^2$, $D_2  = (\ell_{\rm E} - p_2 )^2 + M^2$ and
\begin{align}
N_{\rm E}^{\alpha\beta}  & =  \ell_{\rm E}^{\alpha} \left( C_\chi^2  p_1^\beta + C_{\bar \chi}^2  p_2^\beta \right) \\
&\qquad - C_\chi C_{\bar\chi} \left(-\ell_{\rm E}^{\alpha} p_2^\beta + p_1^\alpha \ell_{\rm E}^{\beta} - p_1^\alpha p_2^\beta \right)\;. \nonumber
\end{align}
Here $p_i$ are the Wick-rotated Euclidean momenta corresponding to $k_i$.
Using Feynman parameters,
\begin{equation}
	\frac{1}{D_0 D_1 D_2} = \int_0^1 dx dy dz \delta(x+y+z-1) \frac{2}{D^3}\;,
\end{equation}
where we define $\Delta = M^2 + y (1-y) p_1^2 + z (1-z) p_2^2 + 2yz p_1 \cdot p_2$,
\begin{align}
	D & = x D_0 + y D_1 + z D_2 \nonumber \\
	  & = \ell_{\rm E}^2 + M^2 + 2\ell_{\rm E}(yp_1 - zp_2) + y p_1^2 + z p_2^2 \nonumber \\
	  & = \left( \ell_{\rm E} + y p_1 - z p_2 \right)^2 + \Delta \;.
\end{align}
In the third line, the $x$ integration has already been performed.
The relevant integral is convergent in the UV, and we are interested in its IR behavior.
Now we can shift $\ell_{\rm E} \to \ell_{\rm E} - (yp_1 - zp_2)$, under which $d^4 \ell_{\rm E} \to d^4 \ell_{\rm E}$ and $D \to \ell_{\rm E}^2 + \Delta$.
\begin{align}
\mathcal T^{\mu\nu}_{12} & = - 4 g^2 M^2 \epsilon^{\mu\nu \alpha \beta} \\
&\quad \times \int_0^1 dy \int_0^{1-y} dz \int \frac{d^4 \ell_{\rm E}}{(2\pi)^4} \frac{\tilde N_{{\rm E},\alpha\beta}}{(\ell_{\rm E}^2 + \Delta)^3}\;, \nonumber
\end{align}
where the shifted numerator is
\begin{equation}
	\tilde N^{\alpha\beta}_{\rm E}
	= -\left[ z C_\chi^2 + y C_{\bar\chi}^2 - (1-y-z) C_\chi C_{\bar\chi} \right] p_1^\alpha p_2^\beta\;.
\end{equation}
Linear terms $\mathcal O(\ell_{\rm E}^\alpha)$ vanish upon integration over $\ell_{\rm E}$.
Thus, only the following integral remains:
\begin{equation}
	\int \frac{d^4 \ell_{\rm E}}{(2\pi)^4} \frac{1}{(\ell_{\rm E}^2 + \Delta)^3} = \frac{1}{32\pi^2} \frac{1}{\Delta}\;.
\end{equation}
In the low-energy limit, where $p_1, p_2, \tilde q \ll M$, one has $\Delta \to M^2$.
The amplitude is given by
\begin{align}
\mathcal T^{\mu\nu}_{12} & = \frac{g^2}{8 \pi^2} \epsilon^{\mu\nu\alpha\beta} k_{1\alpha} k_{2\beta} \int_0^1 dy \int_0^{1-y} dz \\
&\qquad \times \left[ z C_\chi^2 + y C_{\bar\chi}^2 - (1- y - z) C_\chi C_{\bar\chi} \right] \nonumber \\
& = \left( C_\chi^2 + C_{\bar\chi}^2 - C_{\chi} C_{\bar\chi} \right) \frac{g^2}{48 \pi^2}  \epsilon^{\mu\nu\alpha\beta} k_{1\alpha} k_{2\beta}\;. \nonumber
\end{align}
In this limit, exchanging the external gauge-boson lines gives an identical contribution and therefore doubles the amplitude.
The total amplitude is given by
\begin{align}
\mathcal M & =  \left( C_\chi^2 + C_{\bar\chi}^2 - C_{\chi} C_{\bar\chi} \right)  \nonumber \\
&\qquad \times \frac{g^2}{24\pi^2} \varepsilon_\mu(k_1) \varepsilon_\nu(k_2) \epsilon^{\mu\nu\alpha\beta}k_{1\alpha}k_{2\beta}\;.
\end{align}
The corresponding matched EFT operator for $\theta$ is
\begin{align}
\mathcal L_{\rm EFT} & = \frac{g^2}{32\pi^2} c \theta B_{\mu\nu} \tilde B^{\mu\nu}  \\
c & = \frac{1}{3} \left( C_\chi^2 + C_{\bar\chi}^2 - C_{\chi} C_{\bar\chi} \right) \;,
\end{align}
where $\tilde B^{\mu\nu} = \epsilon^{\mu\nu\sigma\rho}B_{\sigma\rho}/2$ and $\epsilon^{0123}=1$.
One can check that, in the Dirac limit where $C_{\bar\chi} = - C_\chi$, this expression reproduces the correct anomaly coefficient.

Because the main text contains both $\theta_1$ and $\theta_2$, two terms appear in Eq.~\eqref{eq:C_BL}.
Note that under the $U(1)_g$ transformation $\theta_i \to \theta_i + \delta \theta_i \alpha_g$, where $\delta \theta_i = q_i$ and $\alpha_g$ is the transformation parameter,
the combination $c_1 \theta_1 + c_2 \theta_2$ is invariant, since
\begin{align*}
  & \delta(c_1 \theta_1 + c_2 \theta_2) = \alpha_g \left( c_1 q_1 + c_2 q_2 \right) \\
= & -\frac{\alpha_g}{3} \left[ \sum_{j=1}^k \left( C_{\chi, j}^3 + C_{\bar \chi,j}^3 \right)  + \sum_{j=1}^l \left(C_{\psi, j}^3 + C_{\bar \psi,j}^3  \right)  \right] = 0\;.
\end{align*}
After applying the mixing relation in Eq.~\eqref{eq:mixing}, one obtains the anomalous coupling in Eq.~\eqref{eq:C_BL}.
The anomaly coefficient $\mathcal C_{B-L}$ is completely fixed once the $U(1)_{B-L}$ charge assignment is determined.

\bibliography{ref}

\end{document}